\documentclass[12pt]{iopart}
\usepackage{bm}
\usepackage{amssymb}
\usepackage{epsfig}
\usepackage{float}
\begin{document}

\def\be{\begin{equation}}
\def\en#1{\label{#1}\end{equation}}
\def\d{\dagger}
\def\bar#1{\overline #1}
\newcommand{\per}{\mathrm{per}}
\newcommand{\eqb}{\begin{eqnarray}}
\newcommand{\eqe}{\end{eqnarray}}
\newcommand{\rd}{\mathrm{d}}
\newcommand{\p}{\partial}
\newcommand{\mA}{\mathcal{A}}
 
\title[Bayesian quantum tomography]{Explicit form of the Bayesian posterior  estimate of a quantum state under the uninformative prior }

\author{  V. S. Shchesnovich}
%\ead{valery@ufabc.edu.br}

\address{Centro de Ci\^encias Naturais e Humanas, Universidade Federal do ABC, Santo Andr\'e, SP, 09210-170 Brazil}

\begin{abstract}
An  analytical solution for the posterior estimate in Bayesian tomography of the  unknown quantum state of an arbitrary quantum system (with a finite-dimensional  Hilbert space)    is found.   First, we derive the   Bayesian  estimate for a pure quantum  state measured  by a set of arbitrary rank-1 POVMs  under  the uninformative (i.e. the unitary invariant or Haar) prior.  The expression for the estimate   involves the matrix permanents of  the Gram matrices with repeated   rows and columns, with the matrix elements being  the scalar products of  vectors  giving the  measurement outcomes. Second, an unknown mixed state  is treated by the Hilbert-Schmidt purification.  In this case,  under the uninformative prior  for the combined pure state,   the posterior estimate of the mixed state of the system  is   expressed  through  the matrix  $\alpha$-permanents of  the Gram matrices  of scalar products of vectors giving the  measurement outcomes. In the mixed case, there is also a free integer  parameter -- the Schmidt number --  which can be  used to optimise the Bayesian reconstruction (for instance, in  case of Schimdt number  being equal to 1, the mixed state estimates reduces to the  pure state estimate). We also discuss the perspectives of  approximate numerical computation and asymptotic analytical evaluation  of the  Bayesian estimate  using the derived formula. 
\end{abstract}

 \maketitle

\section{Introduction}
\label{sec1}

The purpose of  the Quantum Tomography (QT) is  reconstruction of the  quantum state from  the results of measurements \cite{all}.  The  first  QT experiments     were performed  in the early   1990-es \cite{VogelRisken,DAriano,Smithey, Durra}.   Nowadays, most of the  tomographic reconstructions are performed by  using the  Maximum likelihood estimation  (MLE) method  proposed in Ref.  \cite{Hradil}  (see also Refs. \cite{MLE1,MLE2,MLE3,MLE4} for further and recent developments).   In the MLE method one finds  an estimate by maximising   the  likelihood function,   the product of the conditional probabilities of measurement outcomes  given by   the  Born rule.  The MLE estimate   is but a point estimate in the Hilbert space, nevertheless, it  can be supplemented by the maximum-likelihood region, i.e. the  region of largest likelihood among all regions of the same size \cite{MLEreg} (for a similar approach of  confidence regions in QT  see Ref. \cite{CR}).  On the other hand,   the Bayesian  approach  in QT accounts for the available prior information on  the unknown state    by  integrating the likelihood function  with some   prior probability distribution.   It was  pioneered in Ref. \cite{Jones},  where the unitary  invariant prior (Haar measure)  was used  and  the  isotropic measurement scheme was shown to  asymptotically saturate a bound on the extractable information. 

In theory,  convergence of the likelihood function to the Dirac distribution  in the limit of large number of measurements   (see, for instance,  Refs.~\cite{Gnedenko,BayesQT})  causes an effective  decoupling of the  posterior probabilities  from the prior information  and  the  agreement between the MLE and Bayesian estimates (in accordance with the Bernstein - von Mises theorem \cite{BvM}). In practice,  Bayesian approach avoids many pitfalls of the MLE   and is generally believed to be superior to the latter  (see, for instance, Ref. \cite{BK}).    However, the price to pay for the superiority  is  a higher    computational complexity  of the  Bayesian QT  as compared to the MLE method (another subtle point is   the     choice of prior, see for general discussion Ref. \cite{Jaynes}).  A systematic, moreover  analytic,  analysis of  Bayesian   QT for  systems   of $1/2$-spins was carried out  in Ref. \cite{BayesQT}, where the explicit formulae for the  Bayesian estimates   of pure as well as mixed quantum states  were derived for various particular measurement schemes. Though there was no general analytical expression, a table of posterior estimates was presented and growth of  complexity of the formulae was noted.      Numerical   schemes  are being proposed to beat the sheer complexity of the Bayesian QT, for instance based on  the advanced  Monte Carlo methods \cite{MCBQT}. For small systems (e.g. a qubit) it  was   theoretically and experimentally demonstrated \cite{adapBQT1,adaptBQT2} that use of the  adaptive Bayesian QT significantly   improves the    efficiency of reconstruction.   
 
The purpose of the present work is to derive  an explicit analytical formula for  the Bayesian estimate of a quantum state under the uninformative prior.  In the pure state estimation problem (i.e. when it is known that the unknown state is pure)   the  uninformative prior  is  the unitarily invariant (Haar) prior used already in Refs. \cite{Jones, BayesQT}. This is the prior which assigns equal probabilities to equal parts of the projective Hilbert space, i.e. all vectors of the Hilbert space are equally probable. In  the case  of a mixed unknown state  there is no agreement on what prior should  be considered as the uninformative prior, since the  mixed states form  a convex set, not a space. However, similar as in Ref. \cite{BayesQT}, we make a physically relevant assumption that a mixed state is actually a part of the combined pure state of the system and an ancilla (i.e. we use  the Hilbert-Schmidt purification),  where we leave the dimension of the ancilla Hilbert space as an optimisation parameter. This simple trick allows one to introduce into the  Bayesian estimate of the system state an effective dimension of the mixed system state.  The effective dimension depends on the number of significantly nonzero eigenvalues  in the unknown  mixed state. 

The paper is organized as follows.  In section \ref{secQT}  we recall some basics of the Bayesian approach in QT and state our goal. In section \ref{QTpure}
we derive an explicit formula for the Bayesian estimate of a pure quantum state. In section \ref{QTmixed} we apply the  result of section  \ref{QTpure}  to the 
 unknown mixed state via the Hilbert-Schmidt purification with an ancilla system.  In section \ref{secNUM} we review the available methods and insights on the numerical and analytical calculation of the  matrix  $\alpha$-permanents which are integral part of the derived analytical formulae. In section \ref{secCON} we state the main results and problems for future research.

%%%%%%%%%%%%%%%%%%%%%%%%%%%%%%%%%%
\section{Bayesian quantum tomography with the  uninformative prior}
\label{secQT}

Consider the problem of reconstructing  the state of a quantum system with $d$-dimensional Hilbert space $H$. Assume that the state is known to be pure\footnote{The general case of a mixed state   will be considered in a forthcoming work}. The Bayesian  scheme starts with selecting a prior probability distribution, which we can cast in the form $P(\psi)\rd\mu(\psi)$, with $\rd\mu(\psi)$ being the Haar (i.e. unitary invariant) measure in the Hilbert space, i.e.
\be
\rd\mu(\psi) = (d-1)!\delta\left(\sum_{k=1}^d|\psi_k|^2-1\right) \prod_{k=1}^d\frac{\rd\psi^*_k\wedge \rd\psi_k}{2i\pi}
\en{dmu}
where $\psi_k$, $k=1,\ldots,d$, are projections of the state $|\psi\rangle$ on some orthogonal basis in $H$ and the Dirac distribution  $\delta(\ldots)$ accounts for the normalization of the state (here the integration is in the complex plane of each $\psi_k$: ${\rd\psi^*_k\wedge \rd\psi_k}/(2i\pi) = \rd \mathrm{Re}(\psi_k)\rd \mathrm{Im}(\psi_k)/{\pi}$). The Haar measure is normalized such that $\int\rd\mu(\psi) = 1$.  The Haar prior, i.e. $P(\psi)=1$,  assigns  equal probabilities   to equal   areas of the $(2d-1)$-dimensional hypersphere in the real projection of the complex $d$-dimensional space. In other words, all vectors in the Hilbert space are equally probable. Thus, if nothing is known about the true state of the system, except that it is pure, one would naturally set $P(\psi) = 1$  (such prior was first used in Ref. \cite{Jones}). One can rightfully  call this prior uninformative.

An experimentalists is given  an ensemble of systems  which are prepared in the same pure state on  which he/she performs measurements described by a Positive Operator Valued Measure (POVM) or by  a set of POVMs (where an adaptive scheme can be implemented). We assume that all POVMs  from the set have only elements of rank equal to 1\footnote{In the general case one can expand each POVM element in the basis of  eigenvectors and use the below derived formulae.} (but not the orthogonal projectors, in general).  In the  formulae below the vectors from all  POVMs  enter symmetrically, thus it is convenient to use a single index for all   elements  from the set of POVMs. For instance, the resolution of unity  (overcomplete for more that one POVM or single  POVM with nonorthogonal elements) reads
\be
\frac{1}{s}\sum_{k=1}^M|\phi_k\rangle\langle \phi_k| =  I, 
\en{unity}
where $M\ge d$ is the number of all possible outcomes $|\phi_k\rangle$ from all measurements and $s$ is the number of such rank-1 POVMs.    Note that in general   $\langle\phi_k|\phi_l\rangle \ne 0$ for $k\ne l$ and $\langle\phi_k|\phi_k\rangle \ne 1$. 

Let us denote the observed frequency of the result $|\phi_k\rangle$ to be $n_k$ (for each particular POVM the respective frequencies sum to the number of measurements with that particular POVM).  Let us denote the total number of measurements with  all POVMs to be $N$ (thus  $\sum_{k=1}^M n_k = N$). We will label the  sequential outcomes of  measurements    by a Greek index, e.g. $|\phi_{k_\alpha}\rangle$, where $1\le \alpha \le N$ and $1\le k_\alpha\le M$.   The  probability $P(n_1,\ldots,n_M|\psi)$  to obtain the frequencies $n_1,\ldots, n_M$ conditional on the state $|\psi\rangle$ is given by the likelihood function\footnote{Since the order of the results is unimportant, one must use  the multinomial factor at the  likelihood function, but   the factor cancels below due to normalization of the posterior distribution.} following from the Born rule:
\be
P(n_1,\ldots,n_M|\psi) =  \prod_{\alpha=1}^N|\langle\phi_{k_\alpha}|\psi\rangle|^2= \prod_{k=1}^M|\langle\phi_k|\psi\rangle|^{2n_k}.
\en{LF}
The goal is to calculate  the Bayesian estimate  $\rho$ (density matrix) on the system state given by the   formula 
\be
\rho = \int\rd\mu(\psi) \frac{P(n_1,\ldots,n_M|\psi)}{P(n_1,\ldots,n_M)}|\psi\rangle\langle\psi|,
\en{rho}
where the total probability of the outcomes, $P(n_1,\ldots,n_M)$,  reads
\be
P(n_1,\ldots,n_M) = \int\rd\mu(\psi) P(n_1,\ldots,n_M|\psi). 
\en{Ptot}

%%%%%%%%%%%%%%%%%%%%%%%
\section{Analytical formula for Bayesian estimate of a pure state}
\label{QTpure}

Eq. (\ref{rho}) involves integral over a polynomial in $|\psi\rangle$ and $\langle \psi|$ of order $N$. All such integrals can be easily evaluated by using  the following identity
\be
\int\rd\mu(\psi)\left(|\psi\rangle\langle\psi|\right)^{\otimes N} = \frac{S_N}{\Tr{S_N}},\quad S_N \equiv \frac{1}{N!}\sum_{\sigma}P_\sigma,
\en{SN}
where $S_N$ is the projector on the symmetric subspace of  $H^{\otimes N}$, i.e. of  the product of $N$ system  Hilbert spaces ($P_\sigma$ is the unitary operator  representation  of permutation $\sigma$ of the   vectors from the individual  spaces $H$ in the tensor product belonging to $H^{\otimes N}$). Indeed, trivially, the l.h.s. of Eq. (\ref{SN})    is  an operator in the symmetric subspace $H^{\otimes N}$, moreover, it commutes with any permutation operator $P_\sigma$. Thus, it must be proportional to the identity operator in the symmetric subspace, i.e. to $S_N$. By taking the trace of both sides of Eq. (\ref{SN}) and using the Haar measure normalization  one arrives at the denominator on the r.h.s., which is   the number of  basis states in $H^{\otimes N}$:  $\Tr{S_N} = \frac{(N+d-1)!}{N!(d-1)!}$\footnote{One can also verify the identity (\ref{SN}) by direct integration in some basis.}. Now, Eq. (\ref{SN}) trivially leads to the following identity
\eqb
\label{MainId}
&&\int\rd\mu(\psi) \prod_{\alpha=1}^N \langle x_\alpha|\psi\rangle \langle \psi| y_\alpha\rangle = \frac{\left(\langle x_N|\cdot \ldots\cdot \langle x_1|\right)S_N\left( |y_1\rangle\cdot \ldots\cdot |y_N\rangle\right)}{\Tr S_N}\nonumber\\
&& = \frac{(d-1)!}{(N+d-1)!}\per\left(\mathcal{M}\right), \quad\mathcal{M}_{\alpha\beta} \equiv \langle x_\alpha|y_\beta\rangle.
\eqe
Here we have denoted by ``$\per$'' the matrix permanent (see, for instance, Ref.   \cite{Minc}), which is defined for a $N\times N$-dimensional matrix $A$ as follows
\be
\per(A) = \sum_{\sigma}\prod_{i=1}^NA_{i\sigma(i)},
\en{per}
where the sum runs over all permutations $\sigma$ of $N$ column indices of  $A$.  Note that the matrix permanent is invariant under any permutation of rows (or columns) of $A$. 

The identity (\ref{MainId}) allows to evaluate the integral giving the Bayesian estimate  (\ref{rho}). Indeed, we can take an element of $\rho$ between two  basis vectors from $H$, say $\langle e_i|\rho|e_j\rangle$, and use Eq. (\ref{MainId}). We get
\be
\rho =  \frac{1}{N+d} \sum_{i,j=1}^d\frac{|e_i\rangle\per(\mathcal{B}^{(i,j)})\langle e_j|}{\per(\mA)}, 
\en{E20}
where we have introduced two Gram matrices, a $N\times N$-dimensional matrix $\mA$ and a $(N+1)\times(N+1)$-dimensional matrix 
$\mathcal{B}^{(i,j)}$. They read
\be
 \mA_{\alpha,\beta} =  \langle\phi_{k_\alpha}|\phi_{k_\beta}\rangle, \quad \mathcal{B}^{(i,j)} = \left(\begin{array}{cc} \mA & V^{(j)}\\  \left[{V}^{(i)}\right]^\dag & \delta_{i,j} \end{array}\right),\quad V^{(l)}_\alpha =  \langle\phi_{k_\alpha}| e_{l}\rangle. 
 \en{MatAB}

 The expression on the r.h.s. of Eq. (\ref{E20}) can be simplified if we take into account two facts. First, $\mA$ is the matrix of repeated rows and columns of $A_{lk}\equiv \langle\phi_l|\phi_k\rangle$. Denoting  by $ A[n_1,\ldots,n_M|m_1,\ldots,m_M]$ the matrix  with the   $k$-th row of $A$ taken $n_k$ times and the $l$-th column of $A$ taken $m_l$ times, we obtain $\mA =  A[n_1,\ldots,n_M|n_1,\ldots,n_M]$ where $n_k$ is the number of times the $k$-th result of measurement (i.e. $|\phi_k\rangle$) has been  registered. Similarly, the vector-column $V^{(i)}$ consists of repeated scalar products, where each  $\langle e_i|\phi_k\rangle$ enters with the  multiplicity $n_k$. 
By applying  the Laplace expansion (see, for instance, Ref. \cite{Minc})  
to the matrix permanent of $\mathcal{B}_{nm}$ twice, first  with respect to the last column, then with respect to the last row, and using the repeated structure of the column-vector $V^{(i)}$, after some algebra (see details in \ref{A4}) we obtain from Eq.~(\ref{E20}) the following  expression for the posterior density matrix 
\be
\rho =   \frac{1}{N+d}  \left\{ I +  \sum_{k,l=1}^M \frac{n_k n_l \per(\mA(l,k))}{\per(\mA)}|\phi_k\rangle\langle\phi_l|\right\},
\en{rho_res}
where $\mA(l,k) \equiv A[n_1,\ldots,n_l-1,\ldots,n_M|n_1,\ldots,n_k-1,\ldots,n_M]$, i.e. the $(N-1)\times(N-1)$-dimensional  Gram matrix obtained from  $\mA$ by  crossing out one  row  of the scalar products with $\langle \phi_l|$ and one column of the scalar products with $|\phi_k\rangle$.   It is evident that the r.h.s. of Eq. (\ref{rho_res}) is invariant under rescaling of all the vectors by complex scalars $|\phi_k\rangle\to\lambda_k|\phi_k\rangle$ (therefore, one can always consider the vectors to be normalized).    

 The denominator on the r.h.s. of Eq. (\ref{rho_res}) can be expanded by applying  the Laplace    formula (see Eq. (\ref{EA1}) of \ref{A4}) to the matrix permanent and taking into account that matrix $\mA$ consists of repeated rows and columns. We have
\be
\per(\mA) = \sum_{k=1}^M n_k\per\left(\mA(l|k)\right)\langle \phi_l|\phi_k\rangle =  \frac{1}{N}\sum_{k,l=1}^M n_kn_l\per\left(\mA(l|k)\right)\langle \phi_l|\phi_k\rangle. 
\en{den}
For instance, from Eq. (\ref{den}) it is easy to see that the trace of the  posterior density matrix is  indeed equal to 1. Note that the posterior density matrix is such that it assigns \textit{at least } the probability $p(x) = 1/(N+d)$ to any outcome $|x\rangle$ of  future measurement, which is in accordance with what is generally expected on the basis of  Laplace's rule of succession for $d$ excluding outcomes  \cite{Jaynes}  ($|x\rangle$ and the orthogonal complement vectors from the standard basis) (see also  Ref. \cite{BK}). 

The fact that the matrix permanents appear in expression (\ref{MainId}) and,  hence, in  formula (\ref{rho_res}) has  a clear  physical interpretation. Indeed, since our prior is uninformative, all vectors are equally probable before any measurement takes place. Therefore,  the form of the total (i.e. unconditional) posterior probability  of $N$  measurement outcomes    must be invariant with respect to simultaneous  unitary  transformation of all $N$ vector outcomes (equivalent to unitary transformation of the pure system  state on which we integrate with the Haar measure). The only scalars of a set of vectors are their scalar products. The only  permutational invariant scalar (which is  linear in each successive measurement outcome and its  Hermitian conjugate,  as  the Born rule dictates)  is   the  permanent of a Gram matrix of their scalar products, exactly as in Eq. (\ref{MainId}).

Let us consider the special case of a qubit, $d=2$ (detailed analysis of this case was first carried out  in Ref. \cite{BayesQT}, where growing complexity of the analytical formulae as $N$ increases was noted). For any two vectors in the $2$-dimensional Hilbert space we have 
\be
|\phi\rangle\langle\psi| = \frac{1}{2}\left\{\langle\psi|\phi\rangle I  + \sum_{j=1}^3 \langle\psi|\sigma_j|\phi\rangle\sigma_j\right\},
\en{E23}
where $\sigma_j$, $j=1,2,3$, are the Pauli matrices. Inserting   Eq. (\ref{E23}) into the general expression for the posterior density matrix (\ref{rho_res}) and using  expansion (\ref{den}) to simplify  the numerator we obtain  the Bayesian estimate of the qubit  pure state 
\be
\rho = \frac12\left\{ I + \sum_{j=1}^3 v_j\sigma_j\right\},\quad  v_j \equiv \sum_{k,l=1}^M\frac{ n_kn_l\per\left(\mA(l,k)\right)}{(N+2)\per(\mA)}\langle\phi_l|\sigma_j|\phi_k\rangle.
\en{rho2}
Note that the vector $\vec{v}$ is real as it should be. 

Let us consider the simplest example of a von-Neumann POVM, i.e. a single POVM consisting of the orthogonal projectors, $\langle\phi_l|\phi_k\rangle = \delta_{l,k}$. In this special case we get
\[
\per\left(\mA(l,k)\right) = \per\left(\mA(k,k)\right)\delta_{l,k},\quad  \per(\mA) = \prod_{k=1}^d n_k!.
\]
Therefore  Eq. (\ref{rho_res}) delivers  in this case the expected result
\be
\rho = \frac{1}{N+d}\left\{ I + \sum_{k=1}^d n_k |\phi_k\rangle\langle\phi_k| \right\} =\sum_{k=1}^d \frac{n_k+1}{N+d} |\phi_k\rangle\langle\phi_k|,
\en{rhovN}
consistent with Laplace's rule of succession discussed above.

%%%%%%%%%%%%%%%%%%%%%

\section{Analytical formula for Bayesian estimate of a mixed state}
\label{QTmixed}

Now let us consider the general case of a mixed state estimation. As discussed in the Introduction, we lift  the  unknown  mixed state of the system  to a pure state of the system (S) and an ancilla (A) by the  Hilbert-Schmidt purification, i.e. 
\be
\rho^{(S)} = \mathrm{Tr}_A \{|\Psi^{(SA)}\rangle\langle \Psi^{(SA)}| \},
\en{HSrho}
where we denote $d_S$ and $d_A$ the dimensions of the Hilbert space of the system and the ancilla. Any density matrix of the system can be represented in the form (\ref{HSrho}) where $d_A\le d_S$ and coincides with the number of nonzero eigenvalues of $\rho^{(S)}$. 

We consider the measurements performed only on the system, which, as before,  are described by a set of rank-1 POVMs.  We combine the outcomes of all measurements into a single   set with one index, similar as in   Eq. (\ref{unity}). In the combined Hilbert space each  POVM element  is given as 
\be
\Pi_k  = |\phi_k\rangle\langle \phi_k|\otimes I_A = \sum_{j=1}^{d_A}|\phi_k\rangle\langle \phi_k|\otimes |a_j\rangle\langle a_j|,
\en{POVM}
where $|a_j\rangle$, $j=1,\ldots, d_A$ is some unspecified basis in the Hilbert space of the ancilla. In this formulation the total probability of $N$ outcomes
 $|\phi_{k_\alpha}\rangle$, $\alpha = 1,\ldots, N$, where the outcome $|\phi_k\rangle$ appears $n_k$ times,  is given by the following  expression
\be
P(n_1,\ldots, n_M)  = \sum_{j_1=1}^{d_A}\ldots\sum_{j_N=1}^{d_A} \int\rd\mu(\Psi^{(SA)})\prod_{\alpha=1}^N
 |\langle\phi^{(S)}_{k_\alpha},a^{(A)}_{j_\alpha}|\Psi^{(SA)}\rangle|^2,
\en{Ptotal}
where $\mu(\Psi^{(SA)})$ is the uninformative (i.e. the unitary invariant) prior in the combined Hilbert space,   as in  Eq. (\ref{dmu}) and $|\phi^{(S)}_k, a^{(A)}_j\rangle \equiv |\phi_k\rangle |a_j\rangle$.  Using  expression (\ref{MainId}) to evaluate the integral in Eq. (\ref{Ptotal}) we obtain 
\be
P(n_1,\ldots, n_M)  = \sum_{j_1=1}^{d_A}\ldots\sum_{j_N=1}^{d_A} \frac{(d_Sd_A-1)!}{(N+d_Sd_A-1)!}\per\left(\mathcal{C}(j_1,\ldots,j_N) \right) ,
\en{Pres}
where $N\times N$-dimensional  matrix $\mathcal{C}(j_1,\ldots,j_N)$    is a tensor product of two Gram matrices: $\mathcal{C}_{\alpha\beta}(j_1,\ldots,j_N) = \langle\phi_{k_\alpha}|\phi_{k_\beta}\rangle \langle a_{j_\alpha}| a_{j_\beta}\rangle = \langle\phi_{k_\alpha}|\phi_{k_\beta}\rangle \delta_{j_\alpha,j_\beta}$. The summation over the ancilla indices $j_1, \ldots, j_N$  in Eq. (\ref{Pres}) can be carried out by using the following identity 
\be
 \sum_{j_1=1}^{d_A}\ldots\sum_{j_N=1}^{d_A} \prod_{\alpha=1}^N \delta_{j_\alpha,j_\sigma(\alpha)} = d_A^{cyc(\sigma)},
\en{Idsum}
where $cyc(\sigma)$ is the number of disjoint cycles in the cycle decomposition of permutation $\sigma$ (see, for instance, Ref. \cite{Stanley}). 
Here we have taken into account that by Eq. (\ref{per})  $\per(\mathcal{C}(j_1,\ldots,j_N))$ is the sum over all permutations $\sigma$,  where each term has as a  factor the product $\prod_{\alpha=1}^N \delta_{j_\alpha,j_\sigma(\alpha)}$, that each permutation can be uniquely represented as a product of disjoint cycles of the type $j_{\alpha_1}\to j_{\alpha_2} \to \ldots \to j_{\alpha_k} \to j_{\alpha_1}$, and that (due to  the delta-symbols) there is just one free index $j$, running from 1 to $d_A$, corresponding to each cycle. Using Eq. (\ref{Idsum}) into Eq. (\ref{Pres}) we obtain the final expression  for the total probability 
\be
P(n_1,\ldots, n_M)  = \frac{(d_Sd_A-1)!}{(N+d_Sd_A-1)!}\per_{d_A}(\mA),
\en{P}
where   $\mA$ is a Gram matrix with repeated rows and columns, defined as before $ \mA  =A[n_1,\ldots,n_M|n_1,\ldots,n_M] $, and $\per_{\alpha}(\ldots)$ stands for the so-called $\alpha$-permanent (see, for instance, Refs. \cite{VJ,Fr,KM}) which is defined for any complex number $\alpha$. For a $N\times N$-dimensional  matrix $A$ it reads
\be
\per_{\alpha}(A) = \sum_{\sigma}\alpha^{cyc(\sigma)}\prod_{i=1}^N A_{i,\sigma(i)},
\en{perd}
where $cyc(\sigma)$ is  the number of disjoint cycles in the cycle decomposition of permutation $\sigma$. Note that the $\alpha$-permanent is also invariant but under the \textit{simultaneous} permutation of rows and columns of the matrix (in fact, only such a permutational invariance   follows from the obvious permutation  invariance of  the measurement results in the Likelihood function). 

Now we apply a similar strategy to derive the Bayesian estimate of the unknown mixed state as we have used in the pure state estimation in the previous section, i.e. we consider the matrix element $\langle n|\rho^{(S)}|m\rangle$ of the Bayesian estimate for a mixed state 
\be
\rho^{(S)} =  \int\rd\mu(\Psi^{(SA)}) \frac{P(n_1,\ldots,n_M|\Psi^{(SA)})}{P(n_1,\ldots,n_M)}\mathrm{Tr}_A\{|\Psi^{(SA)}\rangle\langle\Psi^{(SA)}|\}
\en{mixest}
and use the result (\ref{P}) for $N+1$ vectors with $\langle \phi_{N+1}| \equiv \langle n|$ and $|\phi_{N+1}\rangle  = |m\rangle$. Using Eq. (\ref{P}) into Eq. (\ref{mixest}) we obtain (here $|e_i\rangle $, $i = 1,\ldots,d_S$, is a basis in the system Hilbert space)
\be
\rho^{(S)}  = \frac{1}{N+d_Sd_A} \sum_{i,j=1}^{d_S}\frac{|e_i\rangle\per_{d_A}(\mathcal{B}^{(i,j)})\langle e_j|}{\per_{d_A}(\mA)},
\en{E20M}
where the matrix  $\mathcal{B}^{(n,m)}$ is  defined  in Eq. (\ref{MatAB}) of the previous section. Eq.~(\ref{E20M}) reduces to Eq. (\ref{E20}) for $d_A=1$, i.e. when the system state is  a pure state. Using the expansion similar to the Laplace's for the $\alpha$-permanent, as shown in \ref{A5}, the mixed state estimate (\ref{E20M}) can be simplified to the following form  
\be
\!\!\!\!\!\!\!\!\!\!\!\!\!\!\!\!\!\!\!\!\!\!\!\!
\rho^{(S)}  = \frac{1}{N+d_Sd_A} \left\{d_AI_S + \frac{\sum_{k,l=1}^M n_k(n_l+[d_A-1]^{\delta_{k,l}})\per_{d_A}(\mA(l,k))|\phi_{k}\rangle\langle\phi_{l}|}{\per_{d_A}(\mA)}\right\},
\en{mixed_res}
 $\mA(l,k)= A[n_1,\ldots,n_l-1,\ldots,n_M|n_1,\ldots,n_k-1,\ldots,n_M]$, i.e. defined  similar as in the pure case, and $I_S$ is the identity operator in the system Hilbert space. 
 
Expression (\ref{mixed_res}) gives a valid density matrix (i.e. the r.h.s. is positive definite Hermitian operator of unit trace) for any positive  integer  value of $d_A$ by derivation. Moreover, the sum in the numerator in the second term in the parenthesis on  the r.h.s. of Eq. (\ref{mixed_res}) is a homogeneous  polynomial of order at most $N$ in $d_A$. 
 
 Let us compare the estimates for the pure and mixed states in the case of  a single von-Neumann POVM. We have in this case by simple computation (similar as in \ref{A5}). 
 \be
 \per_{d_A}(\mA) = n_k d_A  \per_{d_A}(\mA(k,k)).
 \en{perd_rec}
Hence,  by using the orthogonality of the POVM elements from Eqs. (\ref{perd_rec}) and  (\ref{E20M}) we get 
 \be
 \rho^{(S)} =  \frac{d_A}{N+d_Sd_A}\left\{ I_S + \sum_{k=1}^M \frac{n_k}{d_A}|\phi_k\rangle\langle\phi_k|\right\}=\sum_{k=1}^M\frac{n_k+d_A}{N+d_Sd_A}|\phi_k\rangle\langle\phi_k|.
 \en{rho_vN}
Estimate (\ref{rho_vN}) is again consistent with Laplace's rule of succession. Indeed, this is easily seen  by reducing it to  the pure state estimate (\ref{rhovN}) for the same POVM  by  rescaling  the number of measurements as follows $N^\prime \equiv N/d_A$ and  the frequencies as  $n^\prime_k \equiv n_k/d_A$.  Thus, one can interpret the results of each run by a $d_Sd_A$-sided die with each side having a number from 1 to $d_S$ (represented by index $k$ in Eq. (\ref{rho_vN})) and  a color from a set of  $d_A$ colours appearing with equal probability  independently  of the numbers, where we only register the numbers.

 %%%%%%%%%%%%%%%%%%%%%%%
 \section{Perspectives of  numerical calculation of the Bayesian estimate  by  using  the derived formula }
 \label{secNUM}
 
Numerical evaluation  of the $\alpha$-permanent, in general, is  a daunting task. Indeed, it is known that  calculation of the  matrix permanent for a general matrix has the $\#$P-complete level of the computational  complexity \cite{Valiant}. The best known algorithm for the permanent of a general matrix is given by Ryser's method \cite{Ryser} which requires $\mathcal{O}(N^22^N)$ flop operations for the $N\times N$-dimensional  matrix. 

However, this general complexity is significantly reduced by the fact that one has to compute the permanent of  matrices with \textit{repeated} rows and columns. Indeed,  only such matrices   appear in the Bayesian quantum tomography, since for large number of measurements $N$, one will eventually have repetitions of the measurement results, where, generally,  $n_k \gg 1$ for $N\gg d_S$. The expected  complexity of the permanent  for  a matrix with bounded rank $\le R$ is $\mathcal{O}(N^R)$ \cite{Gurvits}.  Moreover,  there is a modified Ryser's algorithm having the  number of necessary flops significantly reduced   down to $\mathcal{O}(N^M)$ for the  case of a $N\times N$-dimensional matrix with repeated columns and/or rows, where $M$ different columns are possible (see, for the  details, Appendix D in Ref. \cite{AsympBS}).

The Ryser's method and its modification apply to the usual matrix permanent. There  is no known generalisation  for the $\alpha$-permanent.   The computational complexity of the latter  is also not yet completely understood. However,  it  is conjectured that the $\alpha$-permanent has the same complexity as the usual permanent for any $\alpha > 0$ \cite{Crane} (note that for $\alpha=-1$ the $\alpha$-permanent is the usual determinant modified by a sign factor). There are   reasons to believe that it is the same as of the usual matrix permanent if $\alpha$ is a positive integer, as in our case.

Another, and perhaps more   crucial,  observation is that the exact values of the coefficients in the Bayesian estimate (\ref{E20M}), expressed through the $\alpha$-permanents, are not needed. One only has to estimate the $\alpha$-permanents  to an error of order $\mathcal{O}(N^{-2})$. Indeed, the zeroth order asymptotic estimate  $\sim \mathcal{O}(1)$ coincides with  the usual Maximum Likelihood estimate due to the above discussed convergence of the two in the limit of large number of measurements (this can be  established also directly by using the   asymptotic approximation of the matrix permanent developed  in Ref. \cite{AsympBS}). Hence, for large number of measurements $N\gg d_S$,  the first  nontrivial contribution of the Bayesian approach   is to  to give the   estimation error for the Maximal Likelihood estimate, which would require computing the $\mathcal{O}(N^{-1})$ term in the asymptotic  approximation of the $\alpha$-permanent.  Indeed, the  Maximal Likelihood  estimate    has   an error of the order at least $\mathcal{O}(N^{-1})$ for large $N\gg d_S$, since this is the order of the  minimal probability assigned to any outcome by Laplace's rule of succession.  Such an error is given by  the first-oder  term in the asymptotic expansion of the derived analytical formula for the Bayesian estimate. 
However, even for the usual permanent, such an asymptotic term  is yet not known  in the explicit form and  one must resort to  numerics.  To date, an efficient method for computing the $\alpha$-permanent, similar as for the usual permanent (outlined in Appendix D in \cite{AsympBS}) is still missing. Hopefully, since the $\alpha$-permanents have many other  important applications in physics and mathematics (see, for instance, Refs. \cite{VJ,Fr,KM,Crane}) such a method will be  found.

\section{Conclusion }
 \label{secCON} 
 
We have  derived the explicit formula for the Bayesian estimate of a mixed quantum state of an arbitrary system with a finite-dimensional Hilbert space, applicable for the uninformative prior distribution of the system state. The uninformative prior used here  is defined as the Haar measure in the  projected Hilbert space of a pure state. For a mixed  system state,  the Hilbert-Schmidt purification to a pure state of a system and an ancilla is used, which  one of the possible physical interpretations of a mixed state.  The explicit formula involves the well-known mathematical object: the  matrix $\alpha$-permanent.   The dimension of the Hilbert space of the  ancilla system is equal to  the number of nonzero eigenvalues of the system density matrix. It can  used  for an  adaptive determination of the effective dimension of the reconstructed density matrix. Perspectives of the numerical calculations based on  the derived formulae are discussed, it is  important that the matrix permanents appearing in the Bayesian QT are of the matrices with repeated rows and columns, for which there is significant reduction of the computational complexity of the matrix permanent (which is, in general, exponentially hard to compute). 
Moreover, the exact coefficients, given by the matrix permanents, are not required, only the first two terms in the asymptotic expansion in $N$ (the number of all measurements) is needed for given the estimate itself and the error of the reconstruction. Therefore, there is significant expectation that in near future one would develop a method, analytical or numerical, for such an approximate evaluation of the matrix $\alpha$-permanents, which would  allow effective computation of the posterior  estimate in the Bayesian QT. We leave this as  a problem  for future research. 
 
\section{Acknowledgements}  This work was supported by  the CNPq    of Brazil.

\appendix

%%%%%%%%%%%%%%%%%%%%%%%%%%%%%%%
\section{Derivation of the formula for posterior density matrix in case of pure state estimation}
\label{A4}

We will use the column or row  expansion formula for the matrix permanent (see, for instance, Ref. \cite{Minc}). For $N\times N$-dimensional matrix $A$  it reads
\be
\per(B) = \sum_{k=1}^N\per(B(k,N)) B_{k,N}, 
\en{EA1}
where $B(l,m)$ is the matrix obtained from matrix $B$ by crossing out the $l$-th row and $m$-th column. Then, using the expansion of Eq. (\ref{EA1})  for the last column (i.e. with respect to $V^{(i)}$)  in the matrix $\mathcal{B}^{(i,j)}$ of Eq. (\ref{MatAB}) of section \ref{secQT} we get
\be
\per(\mathcal{B}^{(i,j)}) = \per\left(\mA\right)\delta_{i,j} + \sum_{\alpha=1}^{N}\per\left(\mathcal{B}^{(i,j)}(\alpha|N+1)\right)\langle\phi_{k_\alpha}|e_i\rangle.
\en{EA2}
Now, using the expansion of Eq. (\ref{EA1}) with respect to the last row in $\mathcal{B}^{(i,j)}(\alpha|N+1)$ (i..e with respect to $\widetilde{V}^{(j)}$) we obtain
\be
\per\left(\mathcal{B}^{(i,j)}(\alpha|N+1)\right) =   \sum_{\beta=1}^{N}\per\left(\mathcal{B}^{(i,j)}(\alpha,N+1|\beta,N+1)\right)\langle{e_j}|\phi_{k_\beta}\rangle.
\en{EA3}
Finally, using that $\mathcal{B}^{(i,j)}(\alpha,N+1|\beta,N+1) = \mA({k_\alpha}|{k_\beta})$ in the notations of section \ref{secQT} and inserting the r.h.s. of Eq. (\ref{EA3}) into Eq. (\ref{EA2}), using that each  vector-row $\langle\phi_l|$ (vector-column $|\phi_k\rangle$)  appears exactly $n_l$ ($n_k$) times  in the summation over $\alpha$ ($\beta$), whereas the coefficient is  $\per\left(\mA(l|k)\right)\langle e_j|\phi_k\rangle\langle\phi_l|e_i\rangle$, we obtain the expression on the r.h.s. of Eq. (\ref{rho_res}) of section \ref{secQT}. 

%%%%%%%%%%%%%%%%%%%%%%%%%%%%%%%%%%%%%%%%%
\section{Derivation of the formula for posterior density matrix in case of mixed  state estimation}
\label{A5}

The derivation is based on the expansion for the $\alpha$-permanent,  similar to the Laplace expansion.The  $d_A$-permanent of $\mathcal{B}^{(i,j)}$  is by definition (\ref{MatAB}) 
\be
\per_{d_A} (\mathcal{B}^{(i,j)}) = \sum_{\sigma} \prod_{\alpha=1}^{N+1}d_A^{cyc(\sigma)} \langle \varphi_{k_\alpha}|\widetilde{\varphi}_{k_\sigma(\alpha)}\rangle,
\en{EB1}
where $|\varphi_{k_\alpha}\rangle = |\widetilde{\varphi}_{k_{\alpha}}\rangle =|\phi_{k_\alpha}\rangle $, $\alpha = 1,\ldots,N$ and $|\varphi_{k_{N+1}}\rangle = |i\rangle$ and $| \widetilde{\varphi}_{k_{N+1}}\rangle  =| j\rangle$. Now, as compared to the similar expansion in Eq. (\ref{EA1})-(\ref{EA2})  for the usual permanent, we need to check  how the number of cycles decomposes for a decomposition of the permutation $\sigma$ over $N+1$ elements into a transposition of $\alpha$-th and $(N+1)$-th elements, $(\alpha,N+1)$, and a permutation $\sigma^\prime$ of the first $N$ elements, i.e.
\be
\sigma = \sigma^{\prime}\cdot (\alpha,N+1).
\en{EB2}
Obviously, if $\alpha = N+1$ (the $N+1$-th element is left in place by $\sigma$) then $cyc(\sigma) = cyc(\sigma^\prime) +1$, since there is one additional $1$-cycle (fixed point $N+1$) in permutation $\sigma$ as compared to the cycle decomposition of $\sigma^\prime$. Otherwise, $\alpha\ne N+1$, both permutations have the same number of cycles, since now $(N+1)$-th element belongs to some cycle of $\sigma^\prime$. Thus we obtain a formula for the decomposition as in Eq. (\ref{EB2})
\be
cyc(\sigma^{\prime}\cdot (\alpha,N+1) ) = cyc(\sigma^{\prime}) + \delta_{\alpha,N+1}. 
\en{EB3}
Using the result (\ref{EB3}) and similar expansion steps as in the derivation of E. (\ref{EA3}) we obtain:
\begin{eqnarray}
\label{EB4}
&&\!\!\!\!\!\!\!\!\per_{d_A} (\mathcal{B}^{(i,j)}) = \per_{d_A}(\mA) d_A\delta_{i,j} +  \sum_{\alpha=1}^{N}\per_{d_A}\left(\mathcal{B}^{(i,j)}(\alpha|N+1)\right)\langle\phi_{k_\alpha}|e_j\rangle
\nonumber\\
&&
 \!\!\!\! \!\!\!\!\!\! = \per_{d_A}(\mA) d_A\delta_{i,j} + \sum_{\alpha,\beta=1}^{N}d_A^{\delta_{\alpha,\beta}}\per_{d_A}\left(\mathcal{B}^{(i,j)}(\alpha,N+1|\beta,N+1)\right)\langle{e_i}|\phi_{k_\beta}\rangle \langle\phi_{k_\alpha}|e_j\rangle.\nonumber\\
&&
\end{eqnarray}
Therefore, we have
\[
\!\!\!\!\!\!\!\!\!\! \!\!\!\!\!\!\!\!\!\!\!\!
\rho^{(S)} = \frac{1}{N+d_Sd_A} \left\{d_AI_S + \frac{\sum_{\alpha,\beta=1}^N d_A^{\delta_{\alpha,\beta}}\per_{d_A}\left(\mathcal{B}^{(i,j)}(\alpha,N+1|\beta,N+1)\right)|\phi_{k_\beta}\rangle\langle\phi_{k_\alpha}|}{\per_{d_A}(\mA)}\right\},
\]
where we have used the permutational invariance of the $\alpha$-permanent with respect to simultaneous permutation of rows and columns. 
Now,   counting the number of equal terms in the numerator and using that for $|\phi_{k_\beta}\rangle = |\phi_k\rangle$ and $|\phi_{k_\alpha}\rangle = |\phi_l\rangle$ we have  $\per_{d_A}\left(\mathcal{B}^{(i,j)}(\alpha,N+1|\beta,N+1)\right) =\per_{d_A}(\mA(l,k))$, we obtain the resulting expression in the form of Eq. (\ref{E20M}).


\vskip 10cm
\begin{thebibliography}{99}

\bibitem{all} M. G. A. Paris and J. \v{R}eh\'{a}\v{c}ek (Eds), \emph{Quantum states estimation}, Lect. Notes Phys. vol. 649 (Springer, Berlin
Heidelberg, 2004).


\bibitem{VogelRisken}  K. Vogel and H. Risken, Phys. Rev. A \textbf{40,} 2847 (1989).

\bibitem{Smithey} D. T. Smithey \textit{et al}, Phys. Rev. Lett. \textbf{70,} 1244 (1993); Phys. Rev. A \textbf{48,} 3159 (1993).

\bibitem{DAriano} G. M. D'Ariano, C. Macchiavello,  and M. G. A. Paris, Phys. Rev. A \textbf{50,}  4298
(1994);  G. M. D'Ariano, U. Leonhardt,  and H. Paul, Phys. Rev. A \textbf{52,} R1801 (1995).

\bibitem{Durra} T. J.  Durra, I. A.  Walmsley, and S. Mukamel,  Phys. Rev. Lett. \textbf{74,}  884 (1995).


\bibitem{Hradil} Z. Hradil, Phys. Rev. A \textbf{55}, R1561 (1997).

\bibitem{MLE1}   J. Fiurasek, Z. Hradil, Phys. Rev. A
\textbf{63}, R020101 (2001).

\bibitem{MLE2} J. \v{R}eh\'{a}\v{c}ek, Z. Hradil, Phys. Rev. Lett. \textbf{88}, 130401 (2002).

\bibitem{MLE3} M. Jezek, J. Fiurasek, Z. Hradil, Phys. Rev. A  \textbf{68}, 012305
(2003).


\bibitem{MLE4} Z. Hradil, D. Mogilevtsev, and J. \v{R}eh\'{a}\v{c}ek,
Phys. Rev. Lett. \textbf{96}, 230401 (2006).


\bibitem{MLEreg} J. Shang \textit{et al}, New J. Phys. \textbf{15}, 123026 (2013).

\bibitem{CR} M. Christandl and R. Renner, Phys. Rev. Lett. \textbf{109}, 120403 (2012)


\bibitem{Jones} K. R. W. Jones, Ann. Phys. \textbf{207,} 140 (1991); J. Phys. A: Math. Gen. \textbf{24,} L1415 (1991);
Phys. Rev. A \textbf{50,} 3682 (1994).



\bibitem{Gnedenko} B. V. Gnedenko,  \textit{The Theory of Probability} (English Translation; Mir Publishers, Moscow, 1978), p. 85.


\bibitem{BayesQT}  V. Bu\u zek \textit{et al}, Ann. Phys. \textbf{266,} 454 (1998). 


\bibitem{BvM}  A. W. van der Vaart,  \textit{Asymptotic Statistics}, p. 138 (Cambridge University Press, 1998). 


\bibitem{BK} R. Blume-Kohout, New J. Phys. \textbf{12},  043034 (2010).

\bibitem{Jaynes} E. T. Jaynes, \textit{Probability Theory: The Logic of Science} (Cambridge University Press, Cambridge 2003).


\bibitem{MCBQT}  F. Husz\'ar and N. M. T. Houlsby, Phys. Rev. A \textbf{85}, 052120 (2012).  


\bibitem{adapBQT1} K. S. Kravtsov \textit{et al}, Phys. Rev. A \textbf{87}, 062122  (2013).


\bibitem{adaptBQT2} D.H. Mahler \textit{et al},  Phys. Rev. Lett.  \textbf{111}, 183601 (2013).

%%%%%%%%%%%%%%%%%%

\bibitem{Minc} H. Minc, \textit{Permanents, Encyclopedia of Mathematics and Its Applications}, Vol. \textbf{6} (Addison-Wesley Publ. Co., Reading, Mass., 1978).

\bibitem{Stanley}   R. P. Stanley, \textit{Enumerative Combinatorics}, 2nd ed., Vol. 1 (Cambridge University Press, 2011). 

\bibitem{VJ} D. Vere-Jones, New Zeal. J. Math. \textbf{26}, 125 (1997). 

\bibitem{Fr} P. Frenkel, Math. Res. Lett. \textbf{17}, 795 (2010). 

\bibitem{KM} S. C. Kou and P. McCullagh, Biometrica \textbf{96} , 635 (2009).  

%%%%%%%%%%%%% not cited yet

\bibitem{Valiant} L. G. Valiant, Theoretical Coput. Sci., \textbf{8} (1979)  189.

 
\bibitem{Ryser} H. Ryser, \textit{Combinatorial Mathematics}, Carus Mathematical Monograph No. 14. (Wiley, 1963).



\bibitem{JSV} M. Jerrum, A. Sinclair, and E. Vigoda, Journal of the ACM, \textbf{51} (2004) 671.


 
\bibitem{Gurvits} L. Gurvits, in Mathematical Foundations of Computer Science,  Lecture Notes in Computer Science, Vol. \textbf{3618} (2005)  447.

\bibitem{AsympBS} V. S. Shchenovich, Int. J. Quant. Inform. \textbf{11}, 1350045 (2013). 


\bibitem{Crane} H. Crane, e-print math-co/1304.1772 (2013). 
 
 


\end{thebibliography}
\end{document}